\documentclass[12pt]{elsarticle}
\usepackage{epsfig}
\usepackage{rotating}
\usepackage{lscape}

\biboptions{sort&compress}

\begin{document}

\begin{flushright}

\end{flushright}

\vskip 0.5cm

\begin{center}

{\large {\bf Search for double beta decay of $^{136}$Ce and
$^{138}$Ce with HPGe gamma detector}}

\vskip 0.5cm

{\bf P.~Belli$^{a}$, R.~Bernabei$^{a,b,}$\footnote{Corresponding
author. {\it E-mail address:} rita.bernabei@roma2.infn.it
(R.~Bernabei).}, R.S.~Boiko$^{c}$, F.~Cappella$^{d,e}$,
R.~Cerulli$^{f}$, F.A.~Danevich$^{c}$, A.~Incicchitti$^{d,e}$, B.N.~Kropivyansky$^{c}$,
M.~Laubenstein$^{f}$, D.V.~Poda$^{c}$, O.G.~Polischuk$^{c,d}$,
V.I.~Tretyak$^{c,d}$}

\vskip 0.3cm

$^{a}${\it INFN sezione Roma ``Tor Vergata'', I-00133 Rome, Italy}

$^{b}${\it Dipartimento di Fisica, Universit$\grave{a}$ di Roma
``Tor Vergata'', I-00133 Rome, Italy}

$^{c}${\it Institute for Nuclear Research, MSP 03680 Kyiv,
Ukraine}

$^{d}${\it INFN sezione Roma, I-00185 Rome, Italy}

$^{e}${\it Dipartimento di Fisica, Universit$\grave{a}$ di Roma
``La Sapienza'', I-00185 Rome, Italy}

$^{f}${\it INFN, Laboratori Nazionali del Gran Sasso, I-67010
Assergi (AQ)}

\end{center}

\noindent
{\bf Abstract}

Search for double $\beta$ decay of $^{136}$Ce and $^{138}$Ce was
realized with 732 g of deeply purified cerium oxide sample
measured over 1900 h with the help of an ultra-low background HPGe
$\gamma$ detector with a volume of 465 cm$^3$ at the STELLA
facility of the Gran Sasso National Laboratories of the INFN
(Italy). New improved half-life limits on double beta processes in
the cerium isotopes were set at the level of $\lim T_{1/2}\sim
10^{17}-10^{18}$~yr; many of them are even two orders of magnitude 
larger than the best previous results.

\section{Introduction}

Double beta ($2\beta$) decay experiments are considered to-date as
an unique way to clarify the nature of the neutrino (Majorana or
Dirac particle), test the lepton number conservation, determine the
absolute scale of neutrino mass and establish the neutrino mass
hierarchy, search for an existence of right-handed admixtures in
the weak interaction and hypothetical Nambu-Goldstone bosons
(Majorons) \cite{Rod11,Ell12,Ver12}. The neutrinoless ($0\nu$)
$2\beta$ decay is forbidden in the Standard Model due to violation
of the lepton number by two units. However, $0\nu2\beta$ decay is
predicted in many Standard Model extensions where the neutrino is
considered as a massive Majorana particle. The two neutrino
($2\nu$) $2\beta$ decay is allowed in the Standard Model, however,
being a second-order process in the weak interactions, this is an
extremely rare decay with the half-lives in the range of
  $T_{1/2}\sim 10^{18}-10^{20}$
yr even for the nuclei with the highest decay probability.

The progress in developments of the experimental techniques during the
last two decades leads to an impressive improvement of sensitivity
to the neutrinoless ($0\nu$) mode of $2\beta^-$ decay up to $\lim
T_{1/2}\sim 10^{23}-10^{25}$ yr \cite{DBD-tab,Giu12,Cre14}, while
the $2\nu2\beta$ decay was detected for 11 nuclides with the
half-lives in the range of $T_{1/2}\sim 10^{18}-10^{24}$ yr
\cite{DBD-tab,Bar10,Saa13}.

The sensitivity of experiments to search for double ``plus'' decay:
double electron capture ($2\varepsilon$), electron capture with
emission of positron ($\varepsilon\beta^{+}$), and double positron
($2\beta^{+}$) decay is much lower. Even the most sensitive
counting experiments give only limits at the level of $\lim
T_{1/2}\sim 10^{18}-10^{21}$ yr
\cite{DBD-tab,Ruk10,Bel11a,Bel12,Bel13}. Recently indications on
the two neutrino double electron capture process in $^{130}$Ba,
$^{132}$Ba \cite{Mesh01,Puj09} and $^{78}$Kr \cite{Gav13} were
reported. At the same time, there is a strong motivation to
develop experimental techniques to search for neutrinoless
$2\varepsilon$ and $\varepsilon\beta^+$ decays, since the
investigation of these processes could refine the mechanism of the
$0\nu2\beta$ decay, if observed: whether it appears mainly due to
the Majorana mass of neutrino or due to the contribution of the
right-handed admixtures in weak interactions \cite{Hir94}.

Cerium contains three potentially $2\beta$ active isotopes:
$^{136}$Ce, $^{138}$Ce and $^{142}$Ce (see Table
\ref{table:ce_2b_isotopes}). The $^{136}$Ce isotope is of particular
interest because of one of the highest energy of decay which
enables even $2\beta^+$ decay allowed only for six nuclei
\cite{DBD-tab}.

\begin{table}[htb]
\caption{Potentially 2$\beta$ active isotopes of cerium.}
\begin{center}
\begin{tabular}{llll}
\hline Transition               & Energy release,           & Isotopic      & Allowed decay              \\
  ~                             & keV                       & abundance, \% \cite{Ber11} & channels \\

\hline
$^{136}$Ce $\to$ $^{136}$Ba     & 2378.53(27) \cite{Kol11}  & 0.185(2)      & $2\varepsilon$, $\varepsilon\beta^+$, $2\beta^+$  \\
~                               & 2378.49(35) \cite{Nes12}  & ~             & ~ \\
~ & ~  & ~ & ~ \\
$^{138}$Ce $\to$ $^{138}$Ba     & 693(10) \cite{Wan12}      & 0.251(2)      & $2\varepsilon$  \\
~ & ~  & ~ & ~ \\
$^{142}$Ce $\to$ $^{142}$Nd     & 1417.2(21) \cite{Wan12}   & 11.114(51)    & $2\beta^-$  \\
\hline
\end{tabular}
\end{center}
  \label{table:ce_2b_isotopes}
\end{table}

\begin{figure*}[htb]
\begin{center}
\resizebox{0.8\textwidth}{!}{\includegraphics{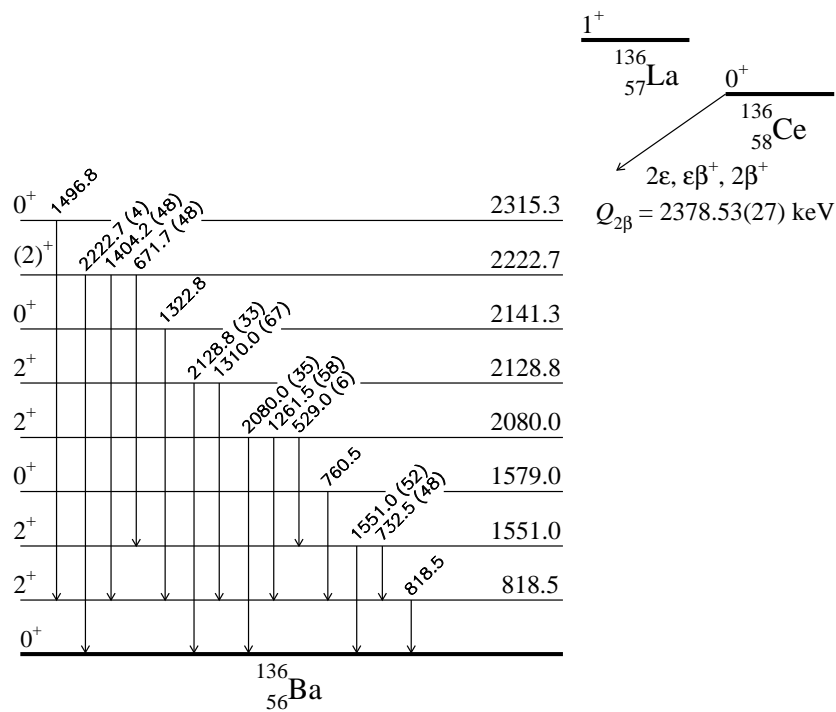}}
\end{center}
\caption{Simplified expected decay scheme of $^{136}$Ce
\cite{Fir98,Son02}. Energies of the excited levels and emitted
$\gamma$ quanta are in keV (relative intensities of $\gamma$
quanta are given in parentheses).}
 \label{fig:136ce-scheme}
\end{figure*}

\begin{figure*}[htb]
\begin{center}
\resizebox{0.5\textwidth}{!}{\includegraphics{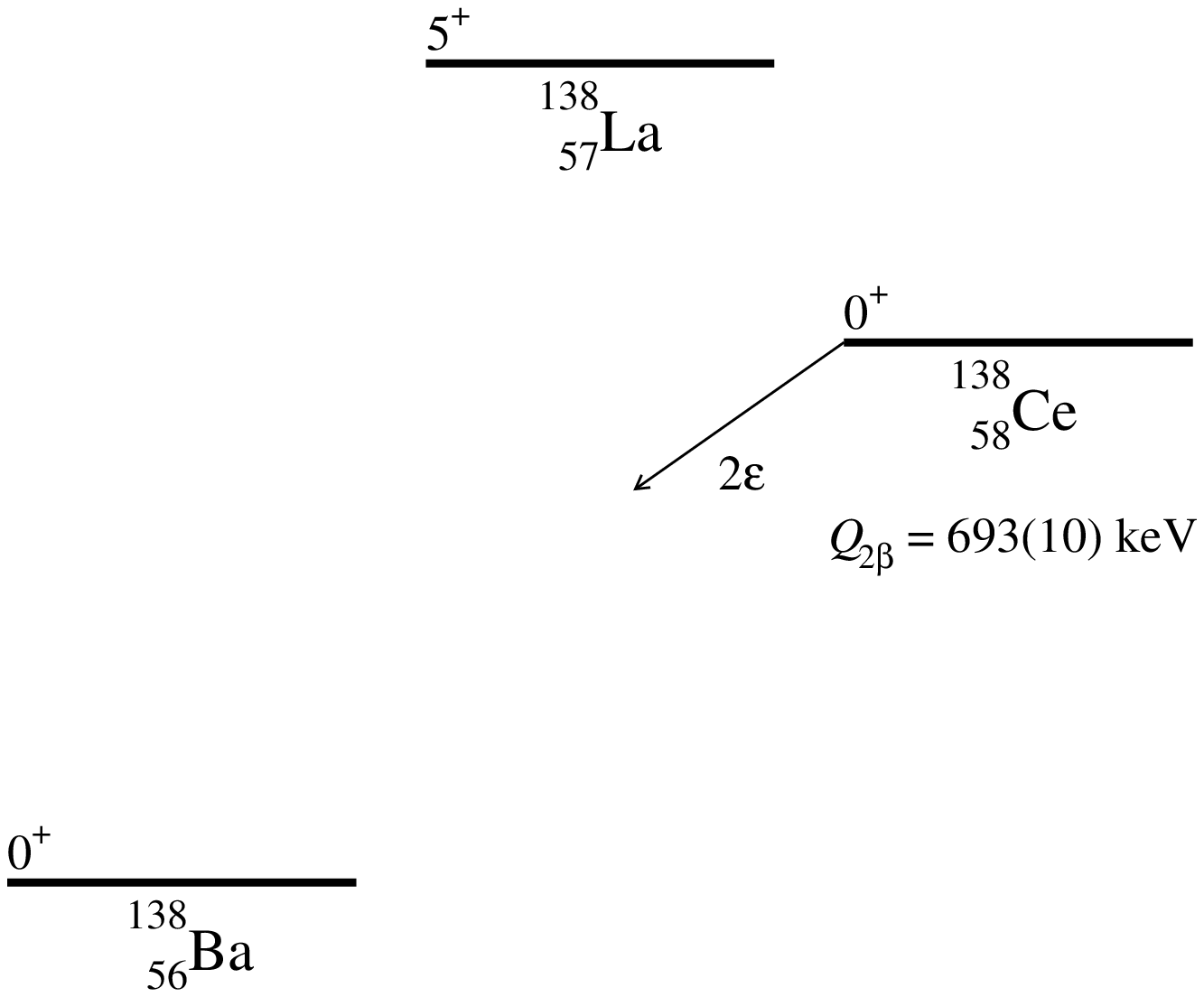}}
\end{center}
\caption{Expected decay scheme of $^{138}$Ce
\cite{Fir98}.}
 \label{fig:138ce-scheme}
\end{figure*}

The searches for double beta decay of the cerium isotopes were
realized in experiments with detectors containing cerium, namely,
gadolinium orthosilicate crystal scintillator doped by cerium
(GSO:Ce) \cite{Dan01}, cerium fluoride (CeF$_3$)
\cite{Ber97,Bel03} and cerium chloride (CeCl$_3$)
\cite{Bel09,Bel11b} crystal scintillators. Double beta processes
in $^{136}$Ce and $^{138}$Ce should be accompanied by emission of
gamma quanta (the decay schemes of $^{136}$Ce and $^{138}$Ce are
presented in Figs. \ref{fig:136ce-scheme} and
\ref{fig:138ce-scheme}, respectively). Therefore, one could apply
gamma spectrometry to search for double beta processes in the
nuclei. The aim of the present work was the search for $2\beta$
processes in $^{136}$Ce and $^{138}$Ce with the help of an
ultra-low background HPGe $\gamma$ detector.

\section{Experiment}

\subsection{Purification of cerium oxide}

Cerium oxide powder of 99.99\% grade was provided by the Stanford
Materials Corporation \cite{Stanford}. A test of the material with
the help of low-background HPGe detector GeBer at the STELLA
facility of the Gran Sasso National Laboratories of the INFN
(Italy) \cite{Arp} showed a considerable contamination of the sample by
potassium, radium, thorium and uranium. The results of the
measurements are presented in Table \ref{table:Ce-cont} (see also
work \cite{Pol13} where preliminary results of the cerium oxide
purification were reported).

To reduce radioactive contamination, the material was purified by
the liquid-liquid extraction method. As a first step, the cerium
oxide was dissolved in a mixture of concentrated nitric and
hydrofluoric acids (the initial amounts of CeO$_2$ and HNO$_3$
were calculated so that to obtain a solution with a 10\%
concentration of Ce(NO$_3$)$_4$ and 5 mol/L of nitric acid):

\begin{center}
 2CeO$_2$ + 4HNO$_3$ + 4HF $\to$ Ce(NO$_3$)$_4$ +
 CeF$_4\downarrow$ + 4H$_2$O.
\end{center}

Then extraction of cerium from the Ce(NO$_3$)$_4$ aqueous solution
was realized using pure tributyl phosphate (TBP) as extragent:

\begin{center}
  Ce(NO$_3$)$_{4,aq}$ + nTBP $\to$ [Ce$\cdot$nTBP](NO$_3$)$_{4,org}$.
\end{center}

\noindent where sub-indexes $aq$ and $org$ denote cerium in
aqueous or organic phase; $n$ is a number of TBP molecules coordinated 
to a Ce$_4^+$ ion. The number $n$ can vary in a wide range from 2 to 5.

Re-extraction of cerium from the organic phase was performed into
low acidic water solution with simultaneous decreasing of the Ce
oxidation level. Hydrogen peroxide was utilized as reducing agent:

\begin{center}
2[Ce$\cdot$nTBP](NO$_3$)$_{4,org}$ + H$_2$O$_2$ $\to$
2Ce(NO$_3$)$_{3,aq}$ + 2HNO$_3$ + O$_2$ + 2nTBP.
\end{center}

Further purification and separation of cerium was performed by
adding of hydrogen peroxide:

\begin{center}
 2Ce(NO$_3)_3$ + 6NH$_3$ + H$_2$O$_2$ + 6H$_2$O $\to$ 2Ce(OH)$_4$ + 6NH$_4$NO$_3$.
\end{center}

The obtained amorphous sediment of Ce(OH)$_4$ was rinsed several
times by ultra-pure water and placed into quartz backers for drying
and annealing to obtain purified cerium oxide.

\begin{table}[htb]
\caption{Radioactive contamination of cerium oxide before and
after purification using liquid-liquid extraction method. Upper
limits are given at 90\% C.L., the uncertainties of the measured
activities are given at $\approx$ 68\% C.L. Radioactive
contamination of CeCl$_3$ and of CeF$_3$ crystal scintillators is given for
comparison.}
\begin{center}
\resizebox{\textwidth}{!}{
\begin{tabular}{lllllll}
\hline
 Chain      & Nuclide     & \multicolumn{5}{c}{Activity (mBq/kg)} \\
\cline{3-7}
~           & ~                     & \multicolumn{2}{c}{CeO$_2$ powder} &  \multicolumn{2}{c}{CeCl$_3$ crystal}    & CeF$_3$ crystal  \\
\cline{3-7}
~           & ~                     & before        & after         & by HPGe       & Scint. mode                   & \cite{Bel03}\\
~           & ~                     & purification  & purification  & \cite{Bel09}  & \cite{Bel11b} \\
\hline
 ~          & $^{40}$K              & 77(28)        & $\leq9$       & $\leq1700$    & --                            & $\leq 330$  \\
 ~          & $^{137}$Cs            & $\leq 3$      & $\leq2$       & $\leq58$      & --                            & -- \\
 ~          & $^{138}$La            & --            & $\leq0.7$     & $680\pm50$    & $862\pm31$                    & $\leq 60$  \\
 ~          & $^{139}$Ce            & --            & $(6\pm1)$    & --           & --                            & --   \\
 ~          & $^{152}$Eu            & --            & $\leq0.5$     & $\leq130$     & --                            & -- \\
 ~          & $^{154}$Eu            & --            & $\leq0.9$     & $\leq60$      & --                            & -- \\
 ~          & $^{176}$Lu            & --            & $\leq0.5$     & $\leq50$      & --                            & $\leq 20$  \\

 $^{232}$Th & $^{228}$Ra            & $850\pm50$    & $53\pm3$      & $\leq210$     & --                            & $890\pm270$ \\
 ~          & $^{228}$Th            & $620\pm30$    & $573\pm17$    & $\leq203$      & $\leq0.16$                   & $1010\pm10$ \\
 ~ & ~ & ~ & ~ & ~ & ~ & ~ \\
 $^{235}$U  & $^{235}$U             & $38\pm10$     & $\leq1.8$     & --            & --                            & $\leq 40$ \\
 ~          & $^{231}$Pa            & --            & $\leq24$      & --            & --                            & $\leq 50$ \\
 ~          & $^{227}$Ac            & --            & $\leq3$       & $\leq740$     & $284\pm2$                     & $\leq 20$ \\
  ~ & ~ & ~ & ~ & ~ & ~ & ~ \\
 $^{238}$U  & $^{238}$U             & $\leq870$     & $\leq40$      & --            & --                            & $\leq 70$ \\
 ~          & $^{226}$Ra            & $11\pm3$      & $\leq1.5$     & $700\pm70$    & $\leq11$                      & $\leq 60$ \\
 ~ & ~ & ~ & ~ & ~ & ~ & \\
\hline
\end{tabular}}
  \label{table:Ce-cont}
\end{center}
\end{table}

\subsection{Low-background measurements}

\begin{figure*}[htb]
\begin{center}
\resizebox{0.8\textwidth}{!}{\includegraphics{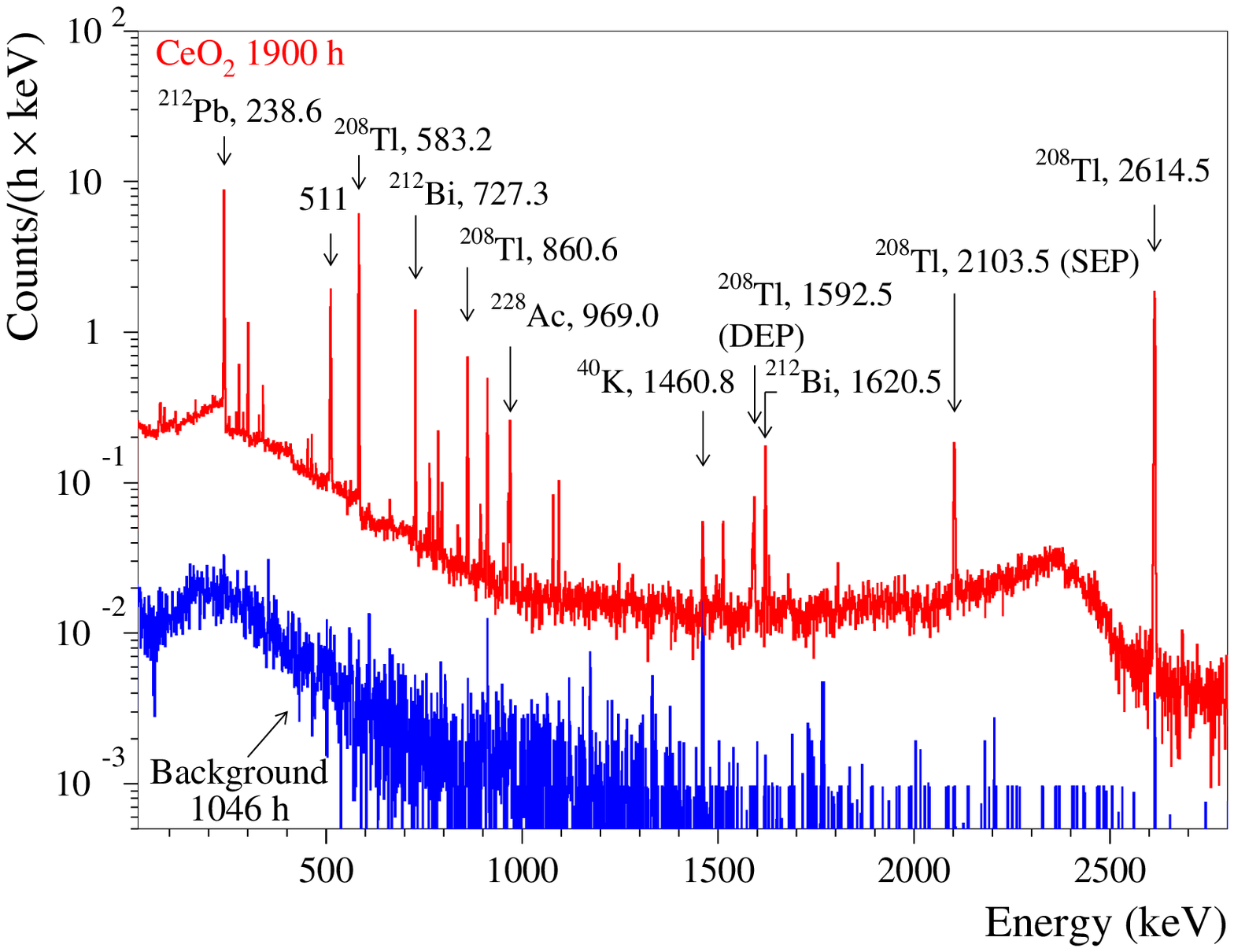}}
\end{center}
\caption{(Color on-line) Energy spectra accumulated with the
CeO$_2$ sample over 1900 h and without sample over 1046 h by the
ultra-low background HPGe $\gamma$ spectrometer. Energy of
$\gamma$ quanta are in keV.}
 \label{fig:E-sp}
\end{figure*}

A sample of the purified cerium oxide with mass 732 g was used in
the experiment carried out in the STELLA low background facility
at the Gran Sasso National Laboratories of the INFN (Italy). The
sample in a thin plastic container was placed on the end-cap of
the ultra-low background HPGe detector GeCris with a volume of 465
cm$^3$. The detector is shielded by low radioactive lead
($\approx25$ cm) and copper ($\approx 10$ cm). The detector energy 
resolution (the full width at the half of maximum, FWHM, keV) can be 
approximated in the energy region of 239--2615 keV by function FWHM =
$\sqrt{1.41(4) + 0.00197(7)\times E_{\gamma}}$, where $E_{\gamma}$ 
is the energy of $\gamma$ quanta in keV.
In particular, FWHM = 2.0 keV at 1332.5 keV. The data
with the sample were accumulated over 1900 h, while the background
spectrum was taken over 1046 h. The energy spectra, normalized to
the time of measurements, are presented in Fig. \ref{fig:E-sp}.

The counting rate of the detector with the cerium sample
substantially exceeds the background of the spectrometer mainly
due to the residual radioactive contamination of the sample by
thorium. Numerous peaks in the spectrum are caused by $\gamma$
quanta of $^{228}$Ac (daughter of $^{232}$Th, in equilibrium with
$^{228}$Ra), $^{212}$Pb, $^{212}$Bi and $^{208}$Tl (daughters of
$^{232}$Th, in equilibrium with $^{228}$Th) with a broken
equilibrium of the $^{232}$Th chain. We have also detected presence 
of cosmogenic (produced by neutrons) $^{139}$Ce in the sample (gamma line with 
energy 165.9 keV). Activities of $^{228}$Ra,
$^{228}$Th and $^{139}$Ce were estimated with the following formula:

\begin{equation}
A = \left( \frac{S_{s}}{t_{s}} - \frac{S_{bg}}{t_{bg}} \right) \cdot \frac{1}{y \cdot \eta \cdot m},
\end{equation}

\noindent where $S_{s}$ ($S_{bg}$) is the area of a peak in the
energy spectrum measured with the CeO$_2$ sample (background);
$t_{s}$ ($t_{bg}$) is the time of the sample (background)
measurements; $y$ is the yield of the corresponding $\gamma$ line
\cite{Fir98}; $\eta$ is the efficiency of the full energy peak
detection; $m$ is the mass of the sample. The detection
efficiencies were calculated by using the EGSnrc code
\cite{EGSnrc} (we have performed calculations also by using GEANT4
package \cite{GEANT4} with the results in good agreement within
3\% with the EGSnrc simulations).

The counting rates in the $\gamma$ peaks of $^{40}$K, $^{137}$Cs,
daughters of $^{235}$U and $^{238}$U in the sample and in the
background (we assume broken equilibrium of the chains):
$^{235}$U, $^{231}$Pa, $^{227}$Ac, $^{238}$U, $^{226}$Ra are
equal inside the statistical uncertainties. Thus, only limits on the
activities of these nuclides were calculated. We have also set
limits on the activity of $^{138}$La, $^{152}$Eu, $^{154}$Eu and
$^{176}$Lu in the sample. The radioactive contamination of the
cerium oxide before and after the purification is summarized in
Table \ref{table:Ce-cont}.

It should be stressed the difference in the contamination of the
cerium oxide and CeCl$_3$ crystal scintillators by thorium. The
activity of $^{228}$Th in the CeCl$_3$ crystal is three orders of
magnitude lower than that in the CeO$_2$ oxide. One can expect
that further purification of cerium from thorium traces is
possible, for instance, by using a chemistry similar to the
applied to produce the CeCl$_3$ crystal scintillators.

\section{Search for double beta decay of cerium}

There are no peculiarities in the spectrum accumulated with the
CeO$_2$ sample which could be attributed to the $2\beta$ processes
in $^{136}$Ce or $^{138}$Ce. Thus, only lower half-life limits can
be estimated by using the following formula:

\begin{equation}
\lim T_{1/2} = N \cdot \eta \cdot t \cdot \ln 2 / \lim S,
\end{equation}

\noindent where $N$ is the number of $^{136}$Ce or $^{138}$Ce
nuclei in the CeO$_2$ sample ($N_{136}$ = 4.74$\times 10^{21}$ and
$N_{138}$ = 6.43$\times 10^{21}$, respectively), $\eta$ is the
detection efficiency, $t$ is the measuring time, and $\lim S$ is
the number of events of the effect searched for which can be
excluded at a given confidence level (C.L., all the limits in the
present study are given at 90\% C.L.). 
Also the detection effciencies of the double beta processes in the cerium 
isotopes were calculated using the EGSnrc code \cite{EGSnrc} and 
GEANT4 code \cite{GEANT4} with
initial kinematics given by the DECAY0 event generator \cite{Decay0}. 
Both codes are again in very good agreement within few percent.

\subsection{Limits on double beta processes in $^{136}$Ce}

In case of the $2\nu2K$ capture in $^{136}$Ce ($^{138}$Ce) a
cascade of X rays and Auger electrons with the individual energies
up to 37.4 keV is expected. The most intensive X ray lines are
expected in the energy region $31.8-37.3$ keV \cite{Fir98}.
However, the detection efficiency for the X rays is too low to
derive reasonable limits on the two neutrino double electron
capture in $^{136}$Ce, e.g., $\eta \approx 1.2\times10^{-8}\%$ and
$\eta \approx 4.8\times 10^{-6}\%$ for energies of gamma quanta
$31.8$ keV and $37.3$ keV, respectively.

The double electron capture in $^{136}$Ce is also allowed to
several excited levels of $^{136}$Ba with energy in the range
$819-2315$ keV. Gamma quanta (cascades of gamma quanta) with
certain energies are expected after the de-excitation of the levels.
For instance, gamma peak with energy 818.5 keV is expected in the
$2\nu2\varepsilon$ decay of $^{136}$Ce to the first excited level
 $2^+$ 818.5 keV of $^{136}$Ba. 

To estimate limits on the
processes, the energy spectrum accumulated with the CeO$_2$ sample
should be fitted in energy regions where the peaks searched for are expected. 
Choice of the energy region for the fit is the main source of the $\lim S$ value uncertainty. 
An energy interval for the fit should be large enough to describe the background precisely. 
From the other side, the interval should not contain peaks (or their areas should be negligible) caused by 
the radioactive contamination of the detector (sample), since weak "hidden" peaks could distort 
background in the region of interest. In some cases, when the region of interest contains a 
background peak near to the peak of effect searched for, it should be included in the model of background. 
Besides, the energy interval should be small enough 
to use whenever possible a simple model to describe the background\footnote{In most of the 
cases background of HPGe detectors in a narrow enough energy interval can be satisfactorily approximated by a linear function.}. 

\begin{figure*}[htb]
\begin{center}
\resizebox{0.65\textwidth}{!}{\includegraphics{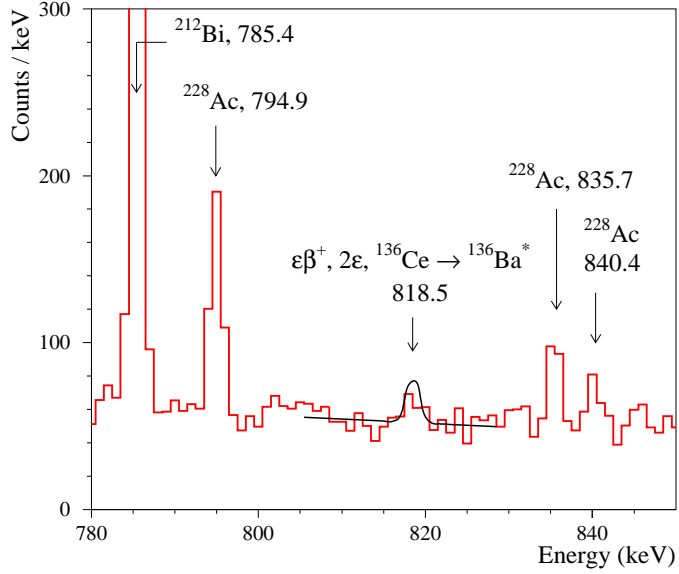}}
\end{center}
\caption{(Color on-line) Energy spectrum measured with the CeO$_2$
sample over 1900 h in the region where the peak at the energy
818.5 keV is expected as a result of de-excitation of excited
levels of $^{136}$Ba due to $2\varepsilon$ or $\varepsilon
\beta^+$ decay of $^{136}$Ce. An excluded peak with energy 818.5
keV and area 53 counts is shown by solid line.} \label{fig:818}
\end{figure*}

The energy interval $798-833$ keV (see Fig. \ref{fig:818}) does not contain any peaks from the radioactive nuclides detected neither in the background energy spectrum nor in the data accumulated with the cerium sample (see subsection 2.2)\footnote{The only significant peak expected in the energy region is 806.2 keV $\gamma$ line of $^{214}$Bi (with an intensity of 1.22\%). However, its contribution, taking into account the limit on $^{226}$Ra contamination in the cerium sample (not to say for a much lower background from $^{214}$Bi) and the detection efficiency, does not exceed 2.5 counts, which is a negligible contribution to the background in the vicinity of the 818.5 keV peak searched for.}.
The fit of the experimental data was performed by a simple model
constructed from a Gaussian function at the energy of 818.5 keV
with the energy resolution FWHM~$=1.74$ keV (the $\gamma$ peak
searched for) and a linear function describing the background. 
The fits were carried out for different energy intervals with start in the range of $798-811$ keV
and end in the range of $825-835$ keV with the step of 1 keV\footnote{It should be stressed, 
the fits give rather stable area of the effect in the range of $25-40$ counts with the error bar in the range of $12-15$ counts.}. 
The best fit by the chi-square method $(\chi^2/$n.d.f. $= 17/20 = 0.85$, where n.d.f. is number of degrees of freedom), 
achieved in the energy interval $805-829$ keV, results in the area of the peak searched for as $S=32\pm13$ counts. 
The Feldman-Cousins procedure \cite{Fel98} gives in this case evidence for the positive effect with area
in the range of $12.0-53.3$ counts at 90\% C.L.; however here we conservatively accept 53 counts
as an effect which can be excluded at 90\% C.L. 
An excluded peak with energy 818.5 keV and area 53 counts is shown in Fig. \ref{fig:818}.
Taking into account the simulated efficiency to detect 818.5 keV
$\gamma$ quanta (2.48\%) we have obtained the following limit on
the $2\nu2\varepsilon$ decay of $^{136}$Ce to the first $2^+$
818.5 keV excited level of $^{136}$Ba:

\begin{center}
 $T_{1/2}^{2\nu2\varepsilon}(^{136}$Ce, g.s.$~\rightarrow~$818.5$)\geq 3.3\times10^{17}$ yr.
\end{center}

Limits on the two neutrino double electron capture to other
excited levels of $^{136}$Ba were obtained by fit of the
experimental spectrum in the energy regions where the most
intensive $\gamma$ peaks are expected with the help of the procedure described above. The obtained half-life
limits are presented in Table \ref{table:Ce_2b_limits}.

In case of $0\nu2\varepsilon$ decay of $^{136}$Ce to the ground state of
$^{136}$Ba, we suppose that only one bremsstrahlung $\gamma$ quantum is
emitted to carry out the transition energy (in addition to X rays
and Auger electrons from de-excitation of atomic shells). 
The energy of the $\gamma$ quantum is expected to be
equal $E_{\gamma} = Q_{2\beta} - E_{b1} - E_{b2}$, where
$E_{b1}$ and $E_{b2}$ are the binding energies of the first and of
the second captured electrons of the atomic shell. The binding
energies on the $K$ and $L_1$, $L_2$, $L_3$ shells in barium atom
are equal to $E_{K} = 37.4$ keV, $E_{L1} = 6.0$ keV, $E_{L2} =
5.6$ keV and $E_{L3} = 5.2$ keV, respectively \cite{Fir98}.
Therefore, the expected energies of the $\gamma$ quanta for the
$0\nu2\varepsilon$ capture in $^{136}$Ce to the ground state of
$^{136}$Ba are: $E_{\gamma} = 2303.7 \pm 0.3$ keV for $0\nu 2K$;
$E_{\gamma} = 2335.5 \pm 0.7$ keV for $0\nu KL$;
$E_{\gamma}=2367.3 \pm 1.1$ keV for $0\nu 2L$. There are no evident
peaks with these energies in the experimental data, 
and we have estimated values of $\lim
T_{1/2}$ by a fit of the data with the procedure described above.

Limits on neutrinoless double electron capture in $^{136}$Ce to
excited levels of $^{136}$Ba were estimated similarly to the two
neutrino double electron capture to the excited levels. The
results are presented in Table \ref{table:Ce_2b_limits}.

\begin{figure*}[!ht]
\begin{center}
\resizebox{0.7\textwidth}{!}{\includegraphics{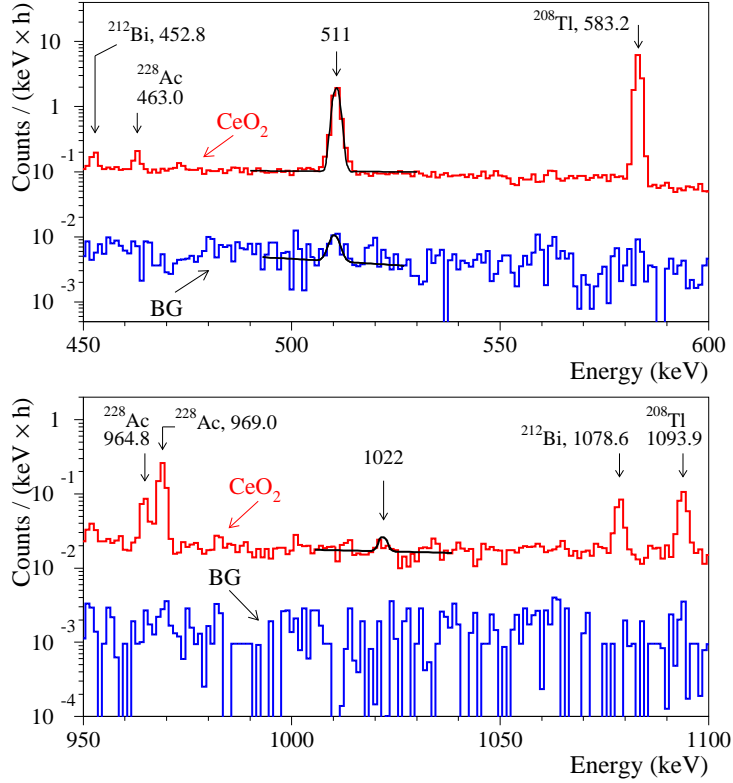}}
\end{center}
\caption{(Color on-line) Part of the energy spectrum accumulated
with the CeO$_2$ sample over 1900 h by the ultra-low background HPGe
$\gamma$ spectrometer (CeO$_2$) in the energy interval $450-600$
keV (upper panel). The energy spectrum accumulated without the sample
over 1046 h (BG) is also reported. Fits of the 511 keV annihilation $\gamma$ peaks are
shown by solid lines. The energy spectra in the energy interval
$950-1100$ keV where a peak with energy 1022 keV due to detection
of two annihilation $\gamma$ quanta is expected (lower panel). The excluded
peak with energy 1022 keV is shown by solid line.}
 \label{fig:511}
\end{figure*}

One (two) positron(s) can be emitted in $\varepsilon\beta^+$
($2\beta^+$) decay of $^{136}$Ce. The annihilation of the
positron(s) will give two (four) 511 keV $\gamma$ quanta leading
to an extra counting rate in an annihilation peak. The energy
spectra accumulated with and without the sample in the energy
interval $450-600$ keV are presented on upper panel of Fig.
\ref{fig:511}. The area of the annihilation peak is $7786\pm532$
counts. 
The area of the peak in the background spectrum
(normalized to the time of measurement with the CeO$_2$ sample) is
$45\pm14$ counts. The excess of events in the data accumulated
with the cerium sample can be explained by the radioactive
contamination of the material by thorium. The decays of $^{228}$Ac
give $22\pm1$ counts and $^{228}$Th daughters provide $8452\pm253$
counts (4974 counts contribute 510.8 keV gamma quanta of
$^{208}$Tl, and 3478 counts come from annihilation $\gamma$ quanta produced
by decays of $^{228}$Th daughters). Therefore, $8474\pm253$ counts
in the annihilation peak can be ascribed to the thorium
contamination of the cerium oxide sample. The difference between
the observed and the expected number of counts (taking also into
account the area of the annihilation peak in the background) $-733\pm
590$ counts gives no evidence for the effect. In accordance with
the Feldman-Cousins procedure, 398 counts can be excluded at 90\%
C.L. Taking into account rather high efficiencies to detect
annihilation $\gamma$ quanta in our experiment (5.58\% and 5.35\% for the
$2\nu\varepsilon\beta^+$ and $0\nu\varepsilon\beta^+$ decay of
$^{136}$Ce to the ground state of $^{136}$Ba, respectively) we
have obtained the following half-life limits:

\begin{center}
 $T_{1/2}^{2\nu\varepsilon\beta^+}(^{136}$Ce, g.s.$~\rightarrow~$g.s.$)\geq 1.0\times10^{17}$
 yr,

 $T_{1/2}^{0\nu\varepsilon\beta^+}(^{136}$Ce, g.s.$~\rightarrow~$g.s.$)\geq 9.6\times10^{16}$ yr.
 \end{center}

To set limits on electron capture with positron emission in
$^{136}$Ce to the $2^+$ 818.5 keV excited level of $^{136}$Ba, the
already obtained $\lim S$ for the expected $\gamma$ peak with
energy 818.5 keV was used. Taking into account the calculated
detection efficiency (1.84\% both for the two neutrino and
neutrinoless process) we have obtained the following limit:

\begin{center}
 $T_{1/2}^{(2\nu,~0\nu)\varepsilon\beta^+}(^{136}$Ce, g.s.$~\rightarrow~818.5$~keV$)\geq 2.5\times10^{17}$ yr.
\end{center}

In case of the double positron decay the highest sensitivity was
reached by analysis of the data in the vicinity of an expected
peak with energy 1022 keV (detection of two annihilation $\gamma$ quanta)
thanks to absence of a peak with this energy in the spectrum
measured with the CeO$_2$ sample (see lower panel of Fig.
\ref{fig:511}).

All the limits on the double beta processes in $^{136}$Ce are
listed in Table \ref{table:Ce_2b_limits}.

\subsection{Search for double electron capture in $^{138}$Ce}

As it was already discussed in section 3.1, the detection
efficiency for X rays expected in two neutrino double electron
capture from $K$ shell of cerium atom is too small to obtain a
competitive limits on the $2\nu2 K$ decay in $^{136}$Ce. The same
conclusion is valid also for the $2\nu2 K$ decay of $^{138}$Ce,
taking into account the same response of the detector to the
process in both cerium isotopes.

\begin{figure}[htb]
\begin{center}
\resizebox{0.7\textwidth}{!}{\includegraphics{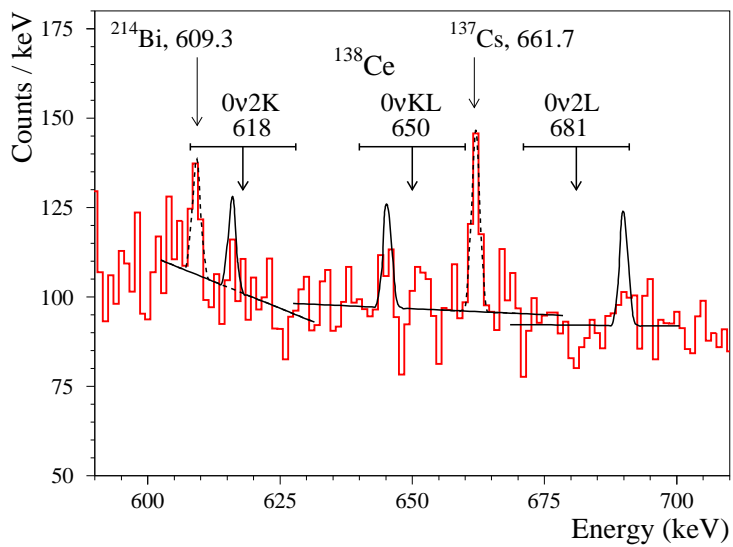}}
\end{center}
\caption{Part of the energy spectrum where peaks from
$0\nu2\varepsilon$ processes ($2K$, $KL$ and $2L$) in $^{138}$Ce
to the ground state of $^{138}$Ba are expected. The excluded peaks
for the processes are shown by solid lines. Peaks of $^{214}$Bi 
with energy 609.3 keV and of $^{137}$Cs with energy 661.7 keV are shown by dashed lines.} \label{fig:138ce_2b}
\end{figure}

To set limits on the $0\nu$ double electron capture in $^{138}$Ce
from $K$ and $L$ shells, one should elaborate wide
energy intervals taking into account the 10 keV uncertainty of the
$Q_{2\beta}$ value for $^{138}$Ce (the energy spectrum with the
regions of interest is presented in Fig. \ref{fig:138ce_2b}). 
Then the peculiarity providing a maximal area of the expected peak within the interval should be accepted to estimate $\lim S$.

To derive a limit on the $0\nu2K$ decay of $^{138}$Ce, a  
model of the background was built from two Gaussians (one to
describe the effect, and the second one to take into account the 
gamma peak with energy 609.3 keV of $^{214}$Bi) plus
first-degree polynomial to describe the background. At the first sight,
it seems that a maximal effect could be associated with the $^{214}$Bi peak. 
To test this assumption, the area of the 609.3 keV peak was bounded within the interval $56\pm13$
counts (area of the peak in the background spectrum normalized on
the time of measurements). The borders of the energy intervals of the fit were varied with
the step of 1 keV within $(595-602)-(630-635)$ keV energy region. The best fit ($\chi^2/$n.d.f. $=21.9/24=0.92$) achieved in the energy interval $600-631$ keV gives the area of the peak
searched for $-12\pm19$ counts (at the energy 610 keV),
therefore we should take $\lim S=20$ counts.
However, there is another peculiarity in the energy interval of interest at the energy $\approx616$ keV.
Thus, we bound energy of the peak searched for within $613-620$ keV (such a restriction is necessary since 
the chi-square algorithm, being applied without the bounds, finds a peak at the previous energy of 610 keV)
and repeated the procedure. 
In this case the best fit (achieved in the energy interval $602-632$ keV, $\chi^2/$n.d.f. $=17.2/23=0.75$) 
gives an area of the effect searched for $18\pm18$ counts, which leads to an estimation of $\lim S=48$ counts 
(the excluded peak is shown in Fig. \ref{fig:138ce_2b}). 
Using the detection efficiency
for $\gamma$ quanta with energy 616 keV (2.74\%) we obtained the
half-life limit:

\begin{center}
 $T_{1/2}^{0\nu2K}(^{138}$Ce, g.s.$~\rightarrow~$g.s.$)\geq 5.5\times10^{17}$ yr.
\end{center}

There are no clear peaks in the energy interval $640-660$ keV
expected in the $0\nu KL$ process in $^{138}$Ce. We have estimated
value of $\lim S$ by fit of the data in the energy intervals
$(625-633)-(669-680)$ keV by a model consisting of two Gaussian functions (one
to describe the peak searched for, and the second one to take into
account the background peak of $^{137}$Cs with energy 661.7
keV\footnote{Despite the 661.7 keV peak does not contribute
directly to the area of the peak searched for, one should include
the peak to describe the background correctly.}) plus polynomial
function (continuous background). The best fit, achieved in the energy interval $627-679$ 
keV, gives area of the effect $31\pm16$ counts at the energy 645 keV, which 
conservatively provides $\lim S=57$ counts.

Finally a fit in the energy interval $668-701$ keV gives the
maximal value $\lim S=63$ counts (at 690 keV) for the $0\nu 2L$
processes in $^{138}$Ce. Taking into account the calculated
efficiencies to detect $\gamma$ quanta with energies 645 keV
(2.69\%) and 690 keV (2.63\%), we set the following limits on the
$0\nu KL$ and $0\nu 2L$ decay of $^{138}$Ce:

\begin{center}

 $T_{1/2}^{0\nu KL}(^{138}$Ce, g.s.$~\rightarrow~$g.s.$)\geq 4.6\times10^{17}$ yr,

 $T_{1/2}^{0\nu 2L}(^{138}$Ce, g.s.$~\rightarrow~$g.s.$)\geq 4.0\times10^{17}$ yr.

\end{center}

\noindent The excluded peaks for the $0\nu$ double electron capture processes 
in $^{138}$Ce are shown in Fig. \ref{fig:138ce_2b}.

The obtained half-life limits on the double beta processes in
$^{136}$Ce and $^{138}$Ce are summarized in Table
\ref{table:Ce_2b_limits} together with the best previous
experimental results and theoretical estimations
\cite{Hir94,Kri11,Aba84,Suh93,Civ98,Rum98,Suh03,Suh12,Bar13}. Current
status of theoretical investigations of neutrinoless 
$\varepsilon\beta^+$ decays is described in \cite{Maa13}.

\clearpage

\begin{landscape}

\begin{table*}[htbp]
\caption{The half-life limits on 2$\beta$ processes in $^{136}$Ce
and $^{138}$Ce together with the best previous limits and
theoretical predictions (the theoretical $T_{1/2}$ values for
$0\nu$ mode are given for $m_{\nu}$ = 1 eV). The energies of the
$\gamma$ lines ($E_{\gamma}$), which were used to set the
$T_{1/2}$ limits, are listed with the corresponding detection
efficiencies ($\eta$) and values of $\lim S$.}
\begin{center}
\resizebox{1.4\textwidth}{!}{
\begin{tabular}{lllllllll}

\hline
 Process  & Decay   & Level of & $E_{\gamma}$  & $\eta$  & $\lim S$ & \multicolumn{2}{c}{Experimental limits,}         & Theoretical estimations, \\
 of decay & mode    & daughter & (keV)         & (\%)    & (cnt)    &  \multicolumn{2}{c}{$T_{1/2}$ (yr) at 90\% C.L.} & $T_{1/2}$ (yr)  \\
\cline{7-8}
 ~        & ~       & nucleus  & ~             & ~       & at 90\%  & Present work            & Best previous       & ~                \\
 ~        & ~       & (keV)    & ~             & ~       & C.L.  & ~                       & results             & ~    \\
 \hline

\multicolumn{9}{l}{$^{136}$Ce $\to$ $^{136}$Ba} \\

  $2K$                 & 2$\nu$ & g.s.    & $31.8-37.3$ & --  & -- & -- & $\geq3.2\times10^{16}$ \cite{Bel11b} & (3.2--9600)$\times10^{18}$ \cite{Rum98,Suh93,Civ98} \\
 $2\varepsilon$       & ~      & $2^+$ 818.5  & 818.5  & 2.48 & 53 & $\geq3.3\times10^{17}$ & $\geq2.5\times10^{15}$ \cite{Bel09} & -- \\
 ~                    & ~      & $2^+$ 1551.0 & 732.5  & 1.07 & 11 & $\geq6.9\times10^{17}$ & $\geq5.4\times10^{15}$ \cite{Bel09} & -- \\
 ~                    & ~      & $0^+$ 1579.0 & 760.5  & 2.19 & 10 & $\geq1.6\times10^{18}$ & $\geq6.3\times10^{15}$ \cite{Bel09} & -- \\
 ~                    & ~      & $2^+$ 2080.0 & 1261.5 & 1.07 & 4.5 & $\geq1.7\times10^{18}$ & $\geq1.3\times10^{15}$ \cite{Bel09} & -- \\
 ~                    & ~      & $2^+$ 2128.8 & 1310.3  & 1.20 & 9.4 & $\geq9.1\times10^{17}$ & $\geq3.2\times10^{15}$ \cite{Bel09} & -- \\
 ~                    & ~      & $0^+$ 2141.3 & 818.5  & 2.17 & 53 & $\geq2.9\times10^{17}$ & $\geq6.1\times10^{15}$ \cite{Bel09} & -- \\
 ~                    & ~      & $(2)^+$ 2222.7 & 671.7 & 1.04 & 10 & $\geq7.4\times10^{17}$ & $\geq5.6\times10^{15}$ \cite{Bel09} & -- \\
 ~                    & ~      & $0^+$ 2315.3 & 818.5 & 2.17 & 53 & $\geq2.9\times10^{17}$ & $\geq5.6\times10^{15}$ \cite{Bel09} & -- \\

 2$K$                 & 0$\nu$ & g.s. & 2303.7 & 1.48 & 23 & $\geq4.6\times10^{17}$ & $\geq3.0\times10^{16}$ \cite{Bel11b} & -- \\
 $KL$                 & 0$\nu$ & g.s. & 2335.5 & 1.48 & 16 & $\geq6.6\times10^{17}$ & $\geq1.4\times10^{15}$ \cite{Bel09} & -- \\
 2$L$                 & 0$\nu$ & g.s. & 2367.3 & 1.45 & 19 & $\geq5.4\times10^{17}$ & $\geq1.1\times10^{15}$ \cite{Bel09} & -- \\

 2$\varepsilon$       & 0$\nu$ & $2^+$ 818.5  & 1485.5  & 1.68 & 4.8 & $\geq2.5\times10^{18}$ & $\geq2.2\times10^{15}$ \cite{Bel09} & -- \\
 ~                    & ~      & $2^+$ 1551.0 & 732.5  & 0.92 & 11 & $\geq6.0\times10^{17}$ & $\geq4.8\times10^{15}$ \cite{Bel09} & -- \\
 ~                    & ~      & $0^+$ 1579.0 & 725.5  & 1.97 & 13 & $\geq1.1\times10^{18}$ & $\geq5.4\times10^{15}$ \cite{Bel09} & -- \\
 ~                    & ~      & $2^+$ 2080.0 & 1261.5  & 0.993 & 4.5 & $\geq1.6\times10^{18}$ & $\geq1.2\times10^{15}$ \cite{Bel09} & -- \\
 ~                    & ~      & $2^+$ 2128.8 & 1310.3  & 1.14 & 9.4 & $\geq8.6\times10^{17}$ & $\geq2.9\times10^{15}$ \cite{Bel09} & -- \\
 ~                    & ~      & $0^+$ 2141.3 & 818.5  &  2.07 & 53 & $\geq2.8\times10^{17}$ & $\geq5.3\times10^{15}$ \cite{Bel09} & -- \\
 ~                    & ~      & $(2)^+$ 2222.7 & 671.7 & 1.03 & 10 & $\geq7.3\times10^{17}$ & $\geq5.4\times10^{15}$ \cite{Bel09} & -- \\
 ~                    & ~      & $0^+$ 2315.3 & 818.5 & 2.16 & 53 & $\geq2.9\times10^{17}$ & $\geq5.4\times10^{15}$ \cite{Bel09} & $1.0\times10^{23}-2.3\times10^{33}$ \cite{Kol11,Kri11,Suh12} \\

 $\varepsilon\beta^+$ & 2$\nu$ & g.s. & 511 & 5.58 & 398 & $\geq1.0\times10^{17}$ & $\geq2.4\times10^{16}$ \cite{Bel11b} & (6.0--9.2)$\times10^{23}$ \cite{Hir94,Rum98} \\
 ~                    & ~      & $2^+$ 818.5  & 818.5  & 1.84 & 53 & $\geq2.5\times10^{17}$ & $\geq2.4\times10^{15}$ \cite{Bel09} & -- \\
 ~                    & 0$\nu$ & g.s. & 511 & 5.35 & 398 & $\geq9.6\times10^{16}$ & $\geq9.0\times10^{16}$ \cite{Bel11b} & $(2.7-4.7)\times10^{26}$ \cite{Hir94,Suh03,Bar13} \\
 ~                    & ~      & $2^+$ 818.5  & 818.5  & 1.84 & 53 & $\geq2.5\times10^{17}$ & $\geq2.4\times10^{15}$ \cite{Bel09} & -- \\

 2$\beta^+$ & 2$\nu$ & g.s. & 1022 & 0.498 & 10 & $\geq3.5\times10^{17}$ & $\geq1.8\times10^{16}$ \cite{Dan01} & 5.2$\times10^{31}$ \cite{Hir94} \\
 ~          & 0$\nu$ & g.s. & 1022 & 0.499 & 10 & $\geq3.6\times10^{17}$ & $\geq6.9\times10^{17}$ \cite{Ber97} & (1.7--2.7)$\times10^{29}$  \cite{Hir94,Suh03,Bar13} \\

 \multicolumn{9}{l}{$^{138}$Ce $\to$ $^{138}$Ba} \\

 2$K$                 & 2$\nu$ & g.s. & $31.8-37.3$ & -- & -- & -- & $\geq4.4\times10^{16}$ \cite{Bel11b} & ~ \\
 2$K$                 & 0$\nu$ & g.s. & $618\pm10$ & 2.74 & 48 & $\geq5.5\times10^{17}$ & $\geq3.6\times10^{16}$ \cite{Bel11b} & $\geq2.1\times10^{26}$ \cite{Aba84} \\
 $KL$                 & 0$\nu$ & g.s. & $650\pm10$ & 2.69 & 57 & $\geq4.6\times10^{17}$ & $\geq4.4\times10^{15}$ \cite{Bel09} & -- \\
 2$L$                 & 0$\nu$ & g.s. & $681\pm10$ & 2.63 & 63 & $\geq4.0\times10^{17}$ & $\geq4.6\times10^{15}$ \cite{Bel09} & -- \\
 \hline

 \end{tabular}
 }
 \end{center}
 \label{table:Ce_2b_limits}
 \end{table*}

\end{landscape}

\clearpage

\section{Conclusions}

Search for double beta processes in $^{136}$Ce and $^{138}$Ce was
realized with the help of an ultra-low background HPGe $\gamma$
detector and a sample of cerium oxide deeply purified by the
liquid-liquid extraction method. Radioactive contamination of the
sample by potassium, radium and uranium was reduced by more than
one order of magnitude. However, the purification procedure was
not efficient enough to remove thorium, which still is present in
the material at the level of 0.6 Bq/kg and remains the main source
of the background.

New improved half-life limits were set on double beta processes in
$^{136}$Ce and $^{138}$Ce at the level of $T_{1/2} \sim
10^{17}-10^{18}$ yr; many of them are even two orders of magnitude 
larger than the best previous results.
At the same time, the sensitivity of the
present experiment is still far from the theoretical predictions,
which are at the level of $T_{1/2} \sim 10^{18}-10^{21}$ yr even
for the most probable two neutrino double electron capture in
$^{136}$Ce (for $2\nu\varepsilon\beta^+$ decay of $^{136}$Ce
$T_{1/2} \sim 10^{24}$ yr), not to say for neutrinoless processes where
theoretical estimations are at the level of $T_{1/2} \sim
10^{26}-10^{29}$ yr (for the effective Majorana neutrino mass
$\langle m_{\nu} \rangle$ = 1 eV).

Further experimental progress can be achieved by deep purification
of cerium from radioactive contamination (mainly by thorium),
using of enriched cerium isotopes, and increase of the experiment
scale.

\section{Acknowledgments}

The group from the Institute for Nuclear Research (Kyiv, Ukraine)
was supported in part by the Space Research Program of the
National Academy of Sciences of Ukraine.

\end{document}